\documentclass[11pt,twoside,a4paper]{article}

\usepackage{simplemargins}
\setleftmargin{2.5cm}
\setrightmargin{2.5cm}
\settopmargin{2.5cm}
\setbottommargin{2.5cm}

\usepackage{graphicx}
\makeatletter
\let\saved@includegraphics\includegraphics
\AtBeginDocument{\let\includegraphics\saved@includegraphics}
\renewenvironment*{figure}{\@float{figure}}{\end@float}
\makeatother
\usepackage[font=small,labelfont=bf]{caption}
\usepackage{times}

\usepackage{lmodern}
\usepackage{textcomp}
\usepackage[latin1]{inputenc}
\usepackage[T1]{fontenc}
\usepackage{amsmath} 
\usepackage{amssymb} 
\usepackage[english]{babel}
\usepackage{sectsty}

\usepackage[normalem]{ulem}
\usepackage{soul}

\usepackage{color}
\usepackage{bm}
\usepackage{comment}
\usepackage{soul}
\usepackage{todonotes}

\usepackage[nomarkers,figuresonly]{endfloat}

\usepackage{float}
\usepackage[normalem]{ulem}

\definecolor{electricpurple}{rgb}{0.75, 0.0, 1.0}
\definecolor{bluegreen}{rgb}{0.0, 0.75, 0.5}

\def\doubleunderline#1{\underline{\underline{#1}}}
\usepackage[square,numbers,sort&compress,comma]{natbib}

\title{Shapely Atoms}

\author
{Andrea Bergamini$^{1,\ast}$, Marco Miniaci$^{1}$, Tommaso Delpero$^{2}$, Gwenael Hannema$^{1}$,\\
	Ivo Leibacher$^{2}$, Armin Zemp$^{1}$}

\begin{document}

\maketitle

%

\small{\noindent $^1$Empa, Laboratory of Acoustics and Noise Control, \"Uberlandstrasse 129, 8600 D\"ubendorf, Switzerland\\
$^2$Empa,Laboratory for Structural Integrity of Energy Systems\"Uberlandstrasse 129, 8600 D\"ubendorf, Switzerland\\
$\ast$Corresponding author: andrea.bergamini@empa.ch}

\begin{abstract}
The study of vibrational properties in engineered periodic structures relies on the early intuitions of Ha\"uy and Boscovich, who regarded crystals as ensembles of periodically arranged mass points interacting via attractive and repulsive forces. Contrary to electromagnetism, where mechanical properties do not couple to the wave propagation mechanism, in elasticity this paradigm inevitably led to low stiffness and high-density materials.
Here, we transcend the Ha\"uy-Boscovich perception, proposing the concept of shaped atoms, which relaxes the link between the mass and inertia of atoms, to achieve unusual dynamic behavior at lower frequencies, leaving the stiffness unaltered. Exploiting tacticity, we successfully demonstrate its feasibility in continuous elastic chiral systems, opening the way to the conception of new mechanisms for wave control, selective wave filtering and vibration isolation.
\end{abstract}

The rational design of periodic structures has recently allowed to attain performances otherwise inaccessible by naturally available materials\textsuperscript{\cite{ZhuNSR2018}}, such as being ultralight and ultrastrong\textsuperscript{\cite{Zheng1373, Meza1322, VIGLIOTTI2012, TancogneAM2018}}, hard to compress yet easy to deform\textsuperscript{\cite{Buckmann2014, LeeAM2012}} or simultaneously exhibiting negative stiffness, Poisson's ratio and/or mass density.\textsuperscript{\cite{Brunet323, Hewage2016}}
This offers unique potential for phonon manipulation\textsuperscript{\cite{MaldovanNature2013,BaiAM2018}} and wave isolation\textsuperscript{\cite{Hussein2014, Delpero2015}} including nonreciprocal\textsuperscript{\cite{popa2014non}} and scattering free\textsuperscript{\cite{MousaviNatComm, MiniaciPhysRevX2018}} propagation, cloaking\textsuperscript{\cite{ZhangPhysRevLett2011}} and frequency band gap nucleation.\textsuperscript{\cite{deymier2013acoustic}}

\begin{figure}
	\centering
	\begin{minipage}[]{1\linewidth}
		{\includegraphics[trim=0in 0in 0.in 0in, clip=true, width=1\textwidth]{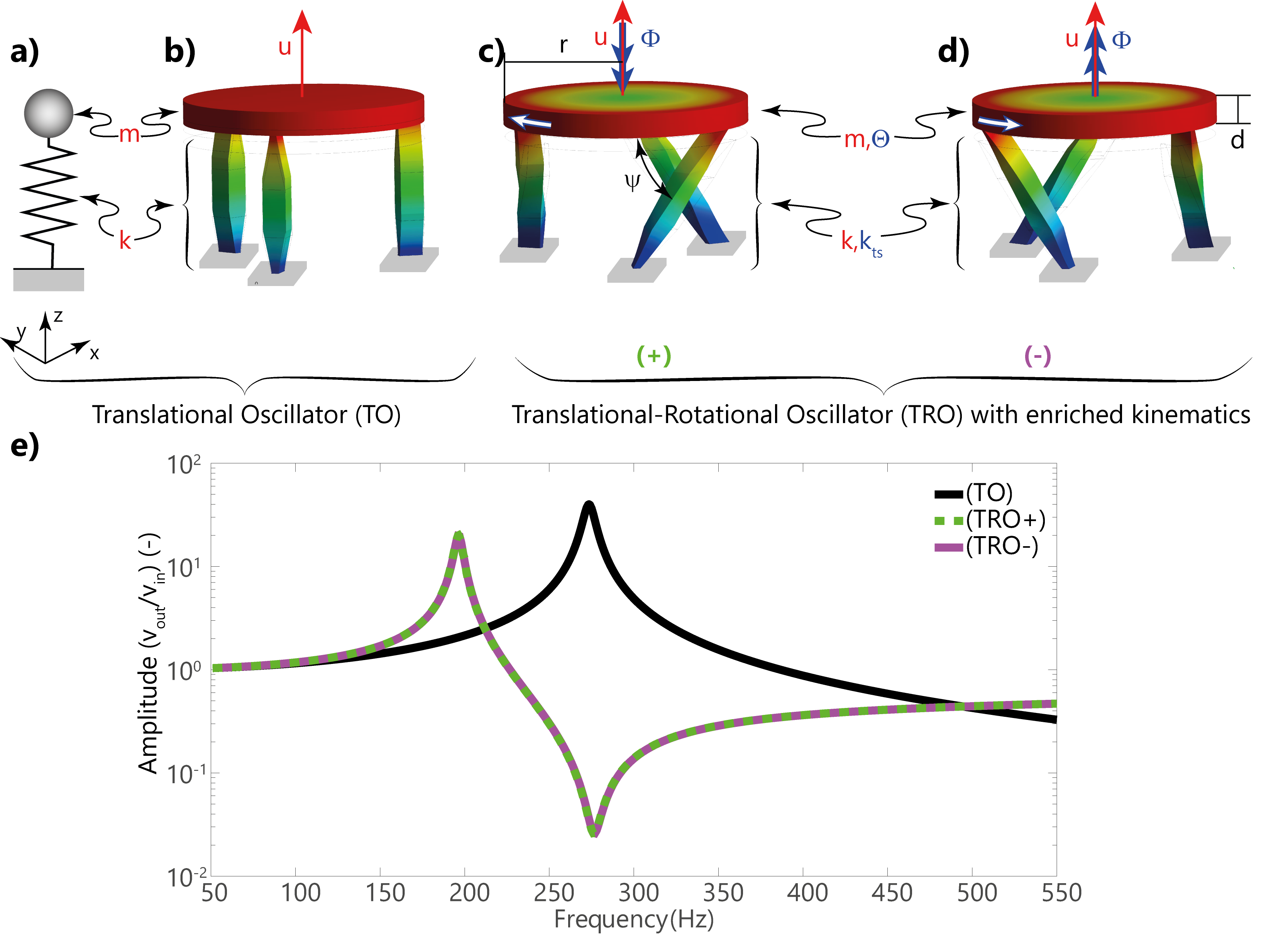}}
	\end{minipage}
	
	\caption{Dynamic behavior of ordinary and enriched kinematics oscillators.
		a)-b), Schematics and three-dimensional model of TO with mass $m$ and axial stiffness $k$. The oscillator can translate in the $z$-direction but cannot rotate, in agreement with the idea that atoms are point masses.
		c)-d),  TROs with enriched kinematics transcending the Ha\"uy-Boscovich model, thanks to their finite extension. Non-centrosymmetric links couple the linear motion ($u$) of the disks along the translational direction ($z$) to their rotation ($\Phi$) about the same axis. The colors in Figs.~\ref{fig1}b-d refer to the magnitude of the displacement at the first resonance of the structures, with blue being the smallest and red the largest total displacement, respectively.
		e), Amplitudes of the vertical displacement as a function of frequency resulting from a numerical forced frequency response analysis. The three structures reported in Figs.~\ref{fig1}b-d are analyzed. Refer to the\textsuperscript{\cite{SM}} for a TRO frequency responce up to higher frequencies.}
	\label{fig1}
\end{figure}

Except for structures including locally resonant elements\textsuperscript{\cite{Liu2000, Casadei2012, Bergamini_AM,Mae1501595}}, so far the tailoring of the dynamics of periodic structures has mainly relied on the Bragg scattering mechanism\textsuperscript{\cite{Brillouin1946}} and the exploration of their vibrational properties on the early intuitions of Ha\"uy and Boscovich, who regarded crystals as periodic\textsuperscript{\cite{Hauy1784}} ensembles of point masses\textsuperscript{\cite{boscovich1922theory}} that interact via attractive and repulsive forces (inter-atomic links). This allowed to conceptualize continuous as well as composite media as a series of discrete mass-spring systems\textsuperscript{\cite{Brillouin1946, Matlack2018NatMat}} in analogy to the atoms and inter-atomic links of a crystal, predicting the essence of the wave propagation as a function of material density $\rho$, stiffness $\mathbb{C}$, and size of a representative unit cell $a$. However, the Ha\"uy-Boscovich model necessarily considers the inter-atomic links as coaxial with the line connecting the atoms. Therefore, an atom can translate but there is no meaning in considering its rotation, leading to the concept of translational oscillator (TO), with a single degree of freedom (DOF) per space direction (Figs.~\ref{fig1}a, b), the motion of which is described by the equation

\begin{equation}
\label{equation_of_motion_SDOF_Oscillator}
\ddot{u} m + u k = 0,
\end{equation}

\noindent being $m$ the mass of the oscillator, $k$ its axial stiffness and $u$ the displacement of the mass. Eq. (\ref{equation_of_motion_SDOF_Oscillator}), applied to a mass-spring chain, implies that the vibrational properties of the TO and consequently the bandgap nucleation, are indissolubly linked to the stiffness and density of the material\textsuperscript{\cite{Brillouin1946}}. The constraint becomes more evident by normalizing the bandgap frequency as $\omega^* = f_{BG} a \sqrt{\frac{\rho}{\mathbb{C}}}$, being $\rho$ the density and $\mathbb{C}$ the stiffness. In the case of a 1D bi-atomic mass spring chain we can show that $1/\pi \le \omega^* \le 2/\pi$\textsuperscript{\cite{SM}}, raising a fundamental theoretical limitation to the conception of structures with simultaneously high stiffness and low density but also small unit cell size and low frequency bandgap, which has indeed been elusive so far.

Here, we transcend the Ha\"uy-Boscovic crystal model by enriching the TO kinematics so as to weaken the link between the mass and the inertia of the atoms by means of non-centrosymmetric links coupling the translational motion of the atoms along one direction to their rotation about the same axis (Figs.~\ref{fig1}c,d), in sharp contrast to ordinary Cauchy media (Figs.~\ref{fig1}a,b), which do not allow for chiral effects\textsuperscript{\cite{Frenzel1072}}.
The non-centrosymmetric architecture, made of elastic elements transferring axial and shear loads as well as bending moments, allows the atoms to rotate, clockwise (Fig.~\ref{fig1}c) or counter-clockwise, (Fig.~\ref{fig1}d) under uni-axial tension, allowing for the conception of a coupled translational-rotational oscillator (TRO).
The twist of the ligaments defines the direction of the atom rotation $\Phi$ with respect to its translation, leading to the definition of two variants of the chiral TRO, $(+)$ and $(-)$. The equation of motion describing the new system has the form of Eq. (\ref{equation_of_motion_SDOF_Oscillator}) but taking into account that now a part of the energy is coupled into a rotary oscillation with the moment of inertia of the disk-shaped atoms

\begin{equation}
\label{equation_of_motion}
\ddot{u}\left(m+\frac{\Theta}{\tan^2(\psi) r^2} \right)+u\left(k+ \frac{k_{ts}}{\tan^2(\psi) r^2}\right)=0,
\end{equation}

\noindent where $k_{ts}$ is the torsional stiffness of the structure, $\psi$ is the angle of the struts between two adjacent disks and $\Theta = \frac{1}{2} m r^2$, assuming that the struts are connected to the external circumference of the disk of radius $r$, as shown in Figs.~\ref{fig1}c,d.  

To unequivocally prove that the introduced rotation, referred to in what follows as \textit{spin}, in analogy to the angular momentum carried by elementary particles in quantum mechanics, allows to relax the restriction imposed by the Ha\"uy-Boscovich model, we performed numerical forced frequency response analyses on the three oscillators reported in Figs.~\ref{fig1}b-d, characterized by the same axial stiffness $k$ and mass $m$. For further details, refer to the supplementary material (SM).\textsuperscript{\cite{SM}} The amplitudes of the total displacement as a function of the exciting frequency are reported in Fig.~\ref{fig1}e for both the achiral TO (black line) and the two chiral TRO with enriched kinematics (superimposed dashed green and solid purple lines). While the harmonic response of the TO is determined by $\sqrt{\frac{k}{m}}$, the introduction of the spin into the unit cell kinematics leads to a considerable reduction of the first rigid atom mode (approximately 30\%), shifting the peak from $273$ Hz in the case of the ordinary TO to $196$ Hz in the case of a TRO. Furthermore, the reconstruction of the mode shapes  clearly confirms the coupling mechanism between torsional and translational oscillation of the shaped atom, as shown in Figs.~\ref{fig1}b-d and detailed in SM.\textsuperscript{\cite{SM}}

The enriched kinematics makes our work distinct from previous studies, as the frequency shift is solely due to the spin introduction and not to any stiffness or mass changes of the unit cell\textsuperscript{\cite{Liu2000, Matlack8386}}.

So far, structured materials with mono-dimensional periodicity have relied on the concatenation of mass-spring oscillators\textsuperscript{\cite{Brillouin1946}}, occasionally including chiral elements\textsuperscript{\cite{BARAVELLI20136562, PalleBigoni, ORTA2019329}}. However, up to now the assembly strategy has been driven by a simple translation of the unit cell, whereas the concatenation of non-centrosymmetric elements into $1D$ objects has been widely studied in the framework of polymer chemistry with respect to substituted poly-olephines, such as polypropylene. The critical effect of the assembly on the physical properties on otherwise chemically identical polymers has been proved\textsuperscript{\cite{natta1955nouvelle, zambelli1968polymerization}}. In this context, the notion of tacticity was introduced as the relative stereochemistry of chiral elements. Inspired by this concept, we show the influence of tacticity on the dynamic behavior of chiral structured periodic media.

\begin{figure}
	\centering
	\begin{minipage}[]{1\linewidth}
		{\includegraphics[trim=0in 0in 0.in 0in, clip=true, width=1\textwidth]{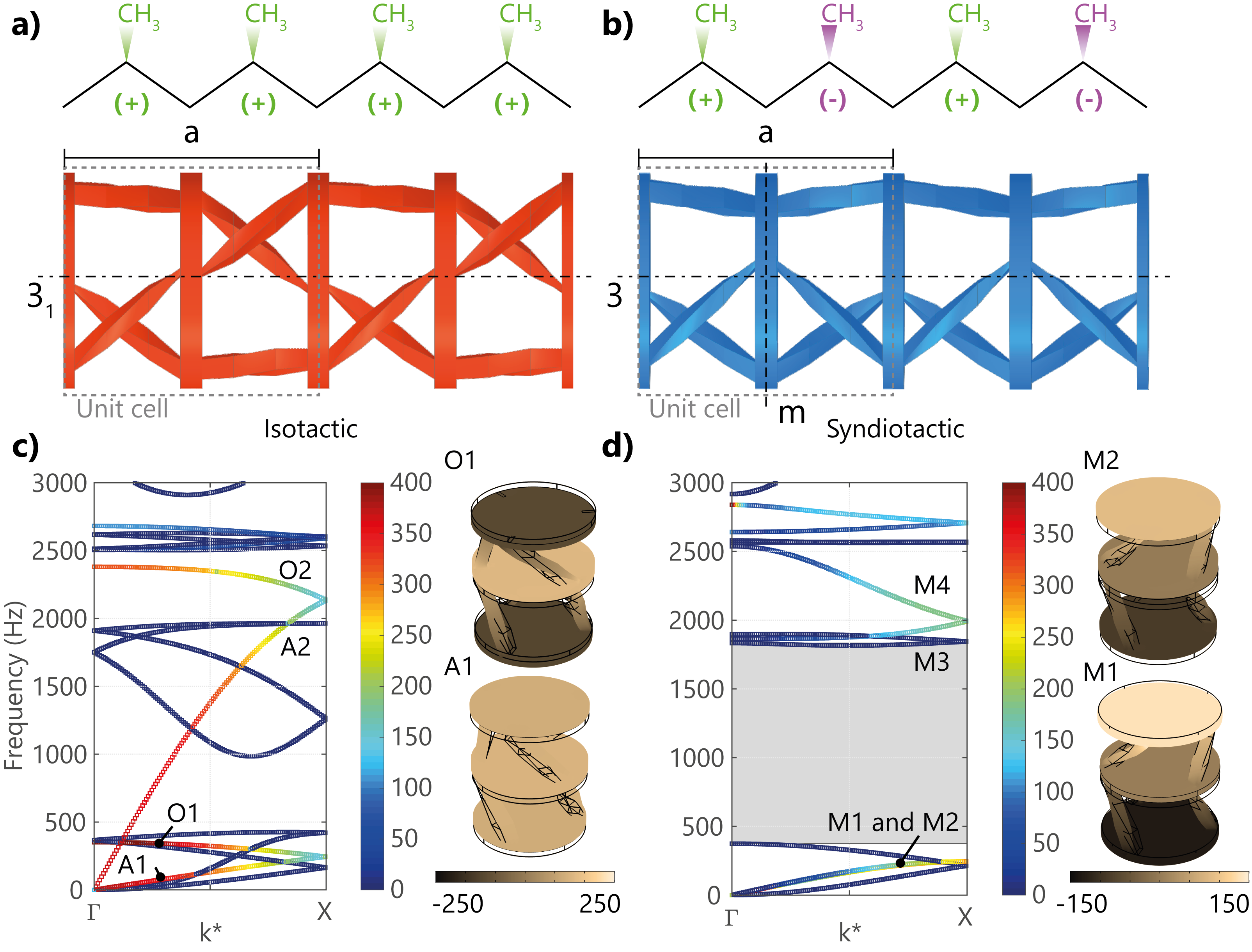}}
	\end{minipage}
	\caption{Influence of the tacticity.
		a)-b), Schematic representation of the concept of tacticity introduced as the relative stereochemistry of chiral elements in the case of the polypropylene (top panels) inspiring the design of two periodic diadic arrays of TROs in isotactic and syndiotactic configuration (bottom panels). The two structures differ for the tacticity of the non-centrosymmetric elements.
		c)-d), Band diagrams for the isotactic and syndiotactic configuration. The dispersion curves are color-coded based on a polarization coefficient that quantifies the rigid motion (rotation +  translation) about the $z$-axis of the finite size atoms. A large bandgap between $374$ Hz and $1816$ Hz in the syndiotactic arrangement is visible.}
	\label{fig2}
\end{figure}

To this end, we considered two phononic crystals (PnCs), consisting of periodic diadic arrays of the TROs  reported in Figs.~\ref{fig1}c,d in isotactic (Fig.~\ref{fig2}a) and syndiotactic (Fig.~\ref{fig2}b) arrangement. In analogy to a polymer (top panels), the \textit{constitution} of the building blocks of the two structures is identical, in both cases periodically arranged in the $z$-direction with the same lattice parameter $a$ and only differ for the \textit{configuration} of the chiral elements (stereoisomers), $(+)/(+)$ and $(+)/(-)$, respectively. Their band diagrams are reported in Figs.~\ref{fig2}c,d and are calculated by imposing Bloch-Floquet periodic boundary conditions over the top and bottom atom surfaces and varying the reduced wavenumber $\mathbf{k}^*=\mathbf{k}_z\cdot\frac{\pi}{a}$ along the $\Gamma - X$ boundary of the first irreducible Brillouin zone (see\textsuperscript{\cite{SM}} for further details).

To better understand the nature of the calculated modes, the dispersion curves are color-coded based on a polarization coefficient $p$\textsuperscript{\cite{MiniaciPhysRevLett}} that quantifies the absolute value of the average $z$-component of the \textit{curl} of the displacement field (i.e., the rotation about the $z$-axis) of the atoms, representative of the enriched rigid body kinematics of the structures reported in Figs.~\ref{fig1}c,d. The polarization factor color bar varies from 0 (blue), indicating that the deformation is localized within the struts, mainly subjected to flexural deformation or as local disk modes, to 400 (red), characterized by a predominantly rigid body motion (rotation+translation) of the disks. This allowed us to identify the modes preeminently involving local deformations (resulting from the continuous nature of the investigated system, blue curves) and those activating the rigid atom rotation in the deformation mechanism of the unit cell (characteristic of the chiral behavior, the remaining curves). Among the latter, we observe that in the isotactic arrangement, the rigid body rotation of its atoms \textit{decreases} as the reduced wavenumber $\mathbf{k}^*$ increases, whereas the syndiotactic structure \textit{increases} the polarization factor $p$ (the color shading of the bands goes from dark blue to light green) as $\mathbf{k}^*$ increases. Inspecting the mode shapes of the two structures, we observe that the tacticity strongly influences the phase of the rotation of the atoms. In the first case (isotactic arrangement), we can recognize the typical behavior of mass-spring chains,\textsuperscript{\cite{Brillouin1946}} where the atoms move in phase along the acoustic branch (A1 in Fig.~\ref{fig2} c) and switch to out of phase after their folding at the $X$ point of the edge of the Brillouin zone (optical mode O1 in Fig.~\ref{fig2} c). However, this is no longer true, in the case of syndiotactic arrangement, where the top and bottom disks are always in opposition of phase leaving the central one still (M1 and M2 in Fig.~\ref{fig2} d). This is a direct consequence of the reversed twist of the ligaments connecting the central disk, implying opposite relative displacement of neighboring disks and torsional stress associated with it. This allows the opening of a large bandgap between $374$ Hz and $1816$ Hz in the syndiotactic arrangement preventing the presence of any propagative mode in this frequency range, contrary to the isotactic case. The key idea is that the syndiotactic arrangement of the atoms leads to alternately stationary (with respect to rotation) and moving disks, whereas in the isotactic one, rotation is accumulated along the $z$-axis, breaking Cauchy media requirements.

To experimentally confirm the above ideas, we prepared two samples, each comprising two unit cells arranged in the iso- and syndiotactic configuration, respectively, as shown in Fig. \ref{fig3} a by means of an additive manufacturing process.
The two structures essentially exhibit the same static stiffness in $z$-direction\textsuperscript{\cite{SM}} and have the same homogenized density ($\rho_h=0.203$ g/cm$^3$).

Next, the transmissibility of the two systems is investigated by scanning laser vibrometry of the two samples (see Fig.~\ref{fig6} for the description of the experimental set-up). The transmissibility is calculated as the ratio of the detected and imposed out of-plane velocity at a scanning point and are presented in Figs.~\ref{fig3}b,c. The experimental results are well supported by finite element numerical models implemented in ANSYS, where the orthotropic material properties were obtained by matching the numerical model response of Fig.~\ref{fig3}c to measured transmissibility data. Nominal data sheet material properties were used as an initial guess. Static tests (see Fig.~\ref{fig4}) were in excellent agreement with these results. The clear drop in the transmission (Fig.~\ref{fig3}c) can be seen as a direct observation of the numerically predicted bandgaps.

In conclusion, we showed that parting from the notion of atoms as point mass elements and embracing the fact that, in PnCs, atoms are bound to have a finite size and can thus be attributed a shape, has demonstrably a dramatic effect on the wave propagation properties of PnCs. This step is a fundamental leap from the idealized model of natural crystals. The introduction of inertia terms, originating from the shape of atoms, in the dispersion equation alone allows to break the otherwise indissoluble link between wave velocity and atomic mass, with obvious practical implications for the phononic materials community. The finite geometry extension of the atoms leads to the definition of a) coupling ratios between kinematic degrees of freedom (linear+rotational) b) an inertial multiplication factor that determines the effect of inertia on wave velocity. Finally, the introduction of the concept of tacticity in the concatenation of chiral elements allows to add an additional layer of architecture that leads to the creation of material variants with substantially differing physical properties, reminiscent of the differentiation between isotactic and syndiotactic polymers.

\begin{figure}
	\centering
	\begin{minipage}[]{1\linewidth}
		{\includegraphics[trim=0in 0in 0.in 0in, clip=true, width=1\textwidth]{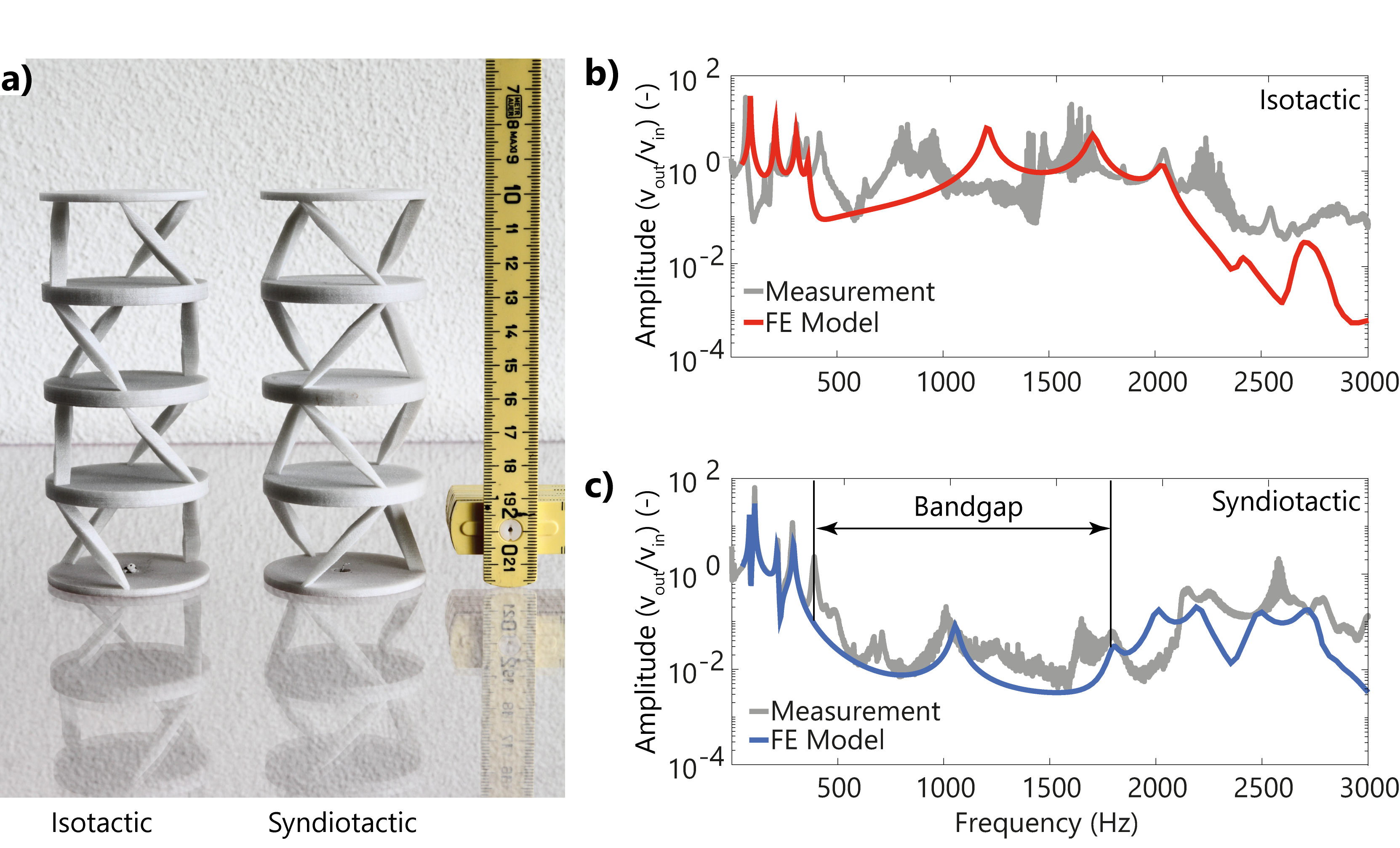}}
	\end{minipage}
	\caption{Measured and calculated transmission spectra.
		a), Photograph of the investigated chiral PnCs. Isotactic sample on the left, syndiotactic on the right, with a cm scale, for reference.
		b), Calculated (red) and measured (grey) transmission spectra of the isotactic crystal.
		c), Calculated (blue) and measured (grey) transmission spectra of the syndiotactic crystal.  The clear drop in the transmission can be seen as a direct observation of the numerically predicted bandgaps.}
	\label{fig3}
\end{figure}




\section*{Acknowledgments}
M.M. has received funding from the European Union's Horizon 2020 research and innovation programme under the Marie Sk{\l}odowska-Curie grant agreement N. 754364. The project that lead to this work was funded by Empa under the Internal Research Call scheme.
The authors wish to thank Mr. Marcel Rees and Mr. Hans Michel of the Mechanical Systems Engineering Laboratory, for their invaluable support in performing the static tests on the PnCs. Also, the authors wish to thank Dr. Bart van Damme and Mr. Kevin Gebhardt for their invaluable support in the execution of the SLDV transmission measurements of the samples. Finally, the authors wish to acknowledge the funding provided by the Empa Board of Directors under the Internal Research Call scheme (IRC2013) that has made this work possible.

\newpage

\section*{Supplementary materials}
\subsection*{Materials and Methods}

\small{\textbf{Simulations}.
Mode shapes showing the different behaviors of the achiral and chiral oscillators and the diagram of the vertical displacement presented in Figs.~\ref{fig1}b-e are calculated by means of ANSYS v19.2 as a result of a numerical forced frequency response analysis. The exciting configuration is the same for the three systems and the force is applied at the center of the disks.

Dispersion diagrams and mode shapes presented in Fig.~\ref{fig2} are computed using Bloch-Floquet theory in full 3D FEM simulations, carried out via the Finite Element solver COMSOL Multiphysics. Full 3D models are implemented to capture all possible wave modes. A material density $\rho_M = 1100 \;\text{kg/m}^3$, based on the measured mass and volume of the PnCs shown in Fig.~\ref{fig3}a, is assumed. The following orthotropic stiffness matrix and Poisson's ratio were assumed for the numerical investigation, respectively: $\doubleunderline{\mathbb{C}}=\left[\begin{array}{ccc}
	3.03& 1.14& 0.86\\
	1.14& 3.03& 0.86\\
	0.86& 0.86& 1.6\\
	\end{array}\right] \cdot10^9$ Pa and $\underline\nu=[\begin{array}{ccc} 0.33&0.33&0.33 \\\end{array}]$.
	
Domains are meshed by means of 4-node quadratic elements of maximum and minimum size $L_{FE}^{max} = 3.3$ mm and $L_{FE}^{min} = 0.2$ mm, respectively, which allowed accurate eigensolutions up to the frequency of interest. Mesh refinement was implemented in proximity of the hinge connections.
The band structures shown in Fig.~\ref{fig2} are obtained assuming periodic conditions along the $z$-direction.
The resulting eigenvalue problem $(\mathbf{K}-\omega^2 \mathbf{M})\mathbf{u} = \mathbf{0}$ is solved by varying the non-dimensional wavevector $\textbf{k}^* = \textbf{k}_z \cdot a$ along the boundary of the irreducible Brillouin zone $ \left[ \Gamma, X \right] $, with $\Gamma \equiv (0, 0) $, $X \equiv (0, \pi/a)$, where $a=59$ mm is the lattice parameter.

The numerical transmission spectra of the isotactic and syndiotactic crystals shown in Fig.~\ref{fig3} are calculated using ANSYS v19.2. The elastic properties of the constituting material (Duraform HST) were estimated from a best match of the numerical and experimental results for the syndiotactic crystal, a good agreement was also found for the isotactic PnC. This approach was used to account for effects on the material properties due to the combination of geometry and manufacturing process. The material properties for the calculations used in Figs. \ref{fig3}b and c were the same as for the calculation of the dispersion curves and the static stiffness (Fig.~\ref{fig4}). 
	
Finally, to better approximate the ideals of stiff mass and massless spring conceptualized in Figs.~\ref{fig1}a,b, the above material properties were modified as follows: the values of the elastic stiffness tensor was multiplied by a factor of 100 for the atoms (approximating the idea of a rigid mass element) and the density of the struts was divided by a factor of 100 (approximating the idea of a massless spring).
Furthermore, the elastic properties of the struts of the TO of Fig.~\ref{fig1}b were divided by a factor of $\approx$100 to match the \textit{static} stiffness of the TROs reported in Figs. \ref{fig1}c,d.
These modifications are justified by the general nature of the considerations presented in Fig.~\ref{fig1}.

\vspace{0.5cm}
\textbf{Experimental measurements}.
The isotactic and syndiotactic PnCs shown in Fig.~\ref{fig3}a are manufactured through selective laser sintering (IRPD AG, St. Gallen, Switzerland).
The specimens were made of Duraform HST, a commercially available mineral fiber reinforced polyamide, with the following \textit{nominal} properties: density $\rho_M = 1200 \; \text{kg/m}^3$, Young modulus $E_{x,y} = 5.5$ GPa, $E_{z} = 3$ GPa, and Poisson ratio $\nu = 0.33$. The geometrical parameters are the following: $a = 59$ mm, $r = 25$ mm, $\psi = \approx \pi/4$, $d = 5$ mm.

The experimental data shown in Figs.~\ref{fig3}b,c were obtained by using a Polytec PSV 400-H scanning laser Doppler vibrometer (SLDV) that measured the out-of-plane velocity of points on a predefined grid (Fig.~\ref{fig6}) over the structure.
The reference input velocity was measured with a Polytec single point LDV (PDV 100) at the edge of the top plate.
Elastic waves were excited through a B\&K  4801 System V by Br\"uel \& Kj\ae r electrodynamic shaker, driven by the PSV 400. The shaker was screwed to the top surface of the plate at the yellow dot shown in Fig.~\ref{fig6}. A linear frequency sweep (with harmonic content ranging from 50 Hz to 3000 Hz) lasting 30 s was used as the excitation signal.

The equivalent static stiffness of the two isotactic and syndiotactic PnCs under rotational free boundary conditions were experimentally verified as well by performing compression tests in a Zwick Z005 universal testing machine. The experimental set-up and the results are shown in Fig.~\ref{fig4}.

\begin{figure*}
\renewcommand{\thefigure}{S\arabic{figure}}
\setcounter{figure}{0}
\centering
\begin{minipage}[]{1\linewidth}
{\includegraphics[trim=0in 0in 0.in 0in, clip=true, width=1\textwidth]{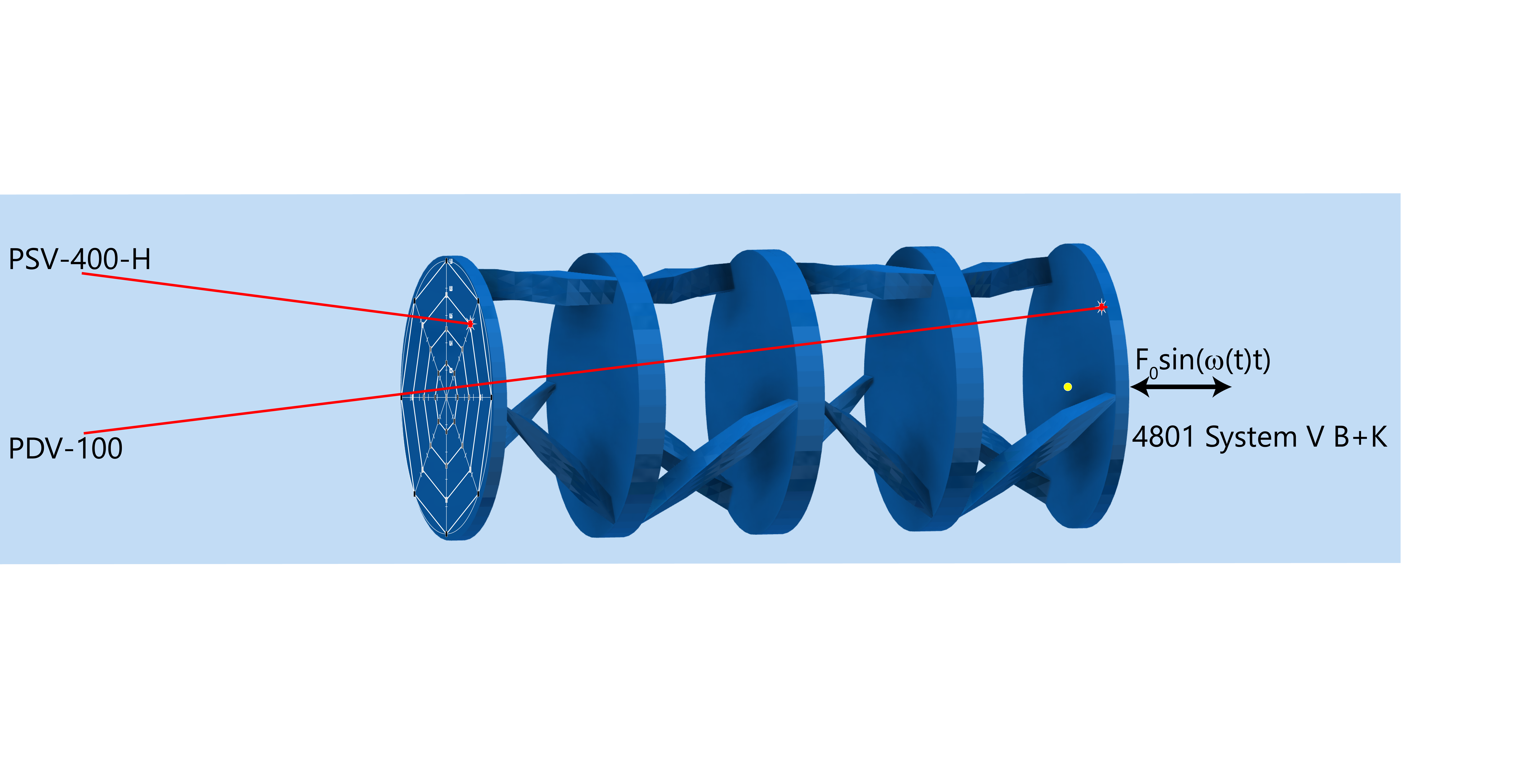}}
\end{minipage}
\caption{Experimental set-up for the measurement of the PnC transmissibility.
The white radar-chart on the left face represents the scanning grid used to determine the average velocity of the face. The input velocity could not be measured at the point at which the mechanical input was applied to avoid excessive parallax errors.
The velocity of the edge of the right plate was deemed a sufficiently good proxy.
However this approximation is responsible for the peak around 1000 Hz visible in transmission function reported in Fig.~\ref{fig3}c, which is due to the fact that the amplitude measured at the edge of the plate becomes extremely small, due to a local mode of the plate, therefore implying a division of the measured output signal by a small value (causing the artificial peak).}
\label{fig6}
\end{figure*}
	
\begin{figure*}
\renewcommand{\thefigure}{S\arabic{figure}}
\setcounter{figure}{1}
\centering
\begin{minipage}[]{1\linewidth}
{\includegraphics[trim=0in 0in 0.in 0in, clip=true, width=1\textwidth]{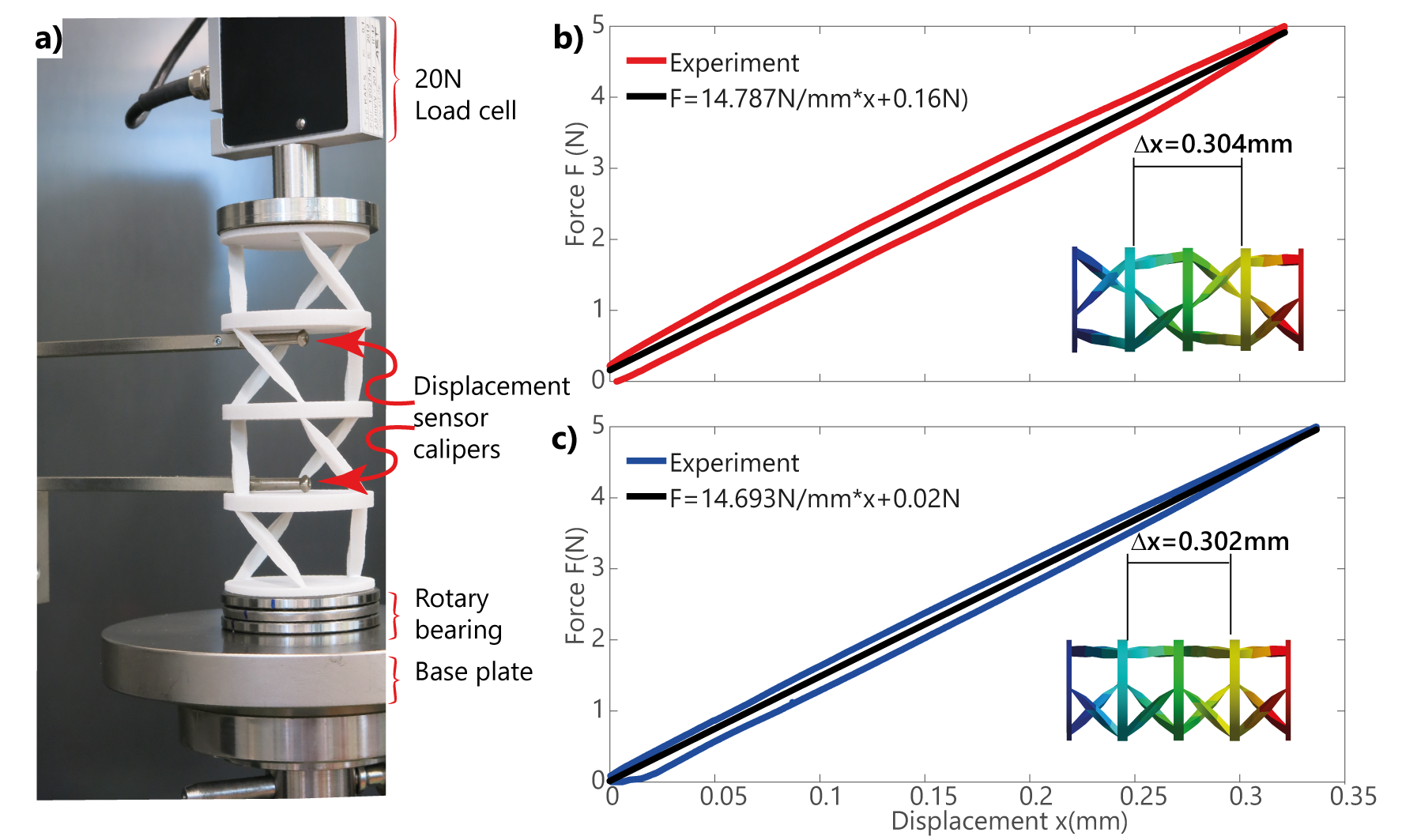}}
\end{minipage}
\caption{Experimental set-up and measurements of the PnC static properties.
a), Experimental set-up for the determination of the PnC static properties. For the sake of simplicity, only the isotactic sample under test is shown.
b), Force as a function of the displacement for the isotactic (red curve) PnC. The inset shows the deformed shape at 5 N compressive load and the relative deformation $\Delta x$ at the measured points.
c), Force as a function of the displacement for the syndiotactic (blue curve) PnC. The inset shows the deformed shape at 5 N compressive load and the relative deformation $\Delta x$ at the measured points.
The results from the linear regression of the data show that the stiffness of the two PnCs is substantially the same.}
\label{fig4}
\end{figure*}

\subsection*{About the limits of $\omega^*$  in a point-mass-spring PnC}
The value of the figure of merit $\omega^*$ is bounded above and below. In the case of a bi-atomic chain, the position of the bottom edge of the bandgap is $\omega_A=\sqrt{\frac{2C}{m_2}}$. The homogenized properties of the 1D-PnC can be expressed in terms of the properties presented in figure \ref{omegastar}. The homogenized stiffness can be expressed as $E_h=\frac{CL}{2A}$, while the homogenized density is $\rho_h=\frac{m_1+m_2}{AL}$. If we now substitute these quantities into the equation expressing $\omega^*$, we have that $\omega^*=\frac{\omega_A}{2\pi}L\sqrt{\frac{\rho_h}{E_h}}=\frac{1}{\pi}\sqrt{1+\frac{m_1}{m_2}}$. The bounds for $\omega^*$ are given by the cases where $m_1\approx m_2$, leading to $\omega^*\rightarrow \frac{2}{\pi}$, or $m_2 \gg m_1$ leading to $\omega^*\rightarrow \frac{1}{\pi}$.
	
\begin{figure*}
\renewcommand{\thefigure}{S\arabic{figure}}
\setcounter{figure}{2}
\begin{center}
{\includegraphics[trim=0in 0in 0.in 0in, clip=true, width=.7\textwidth]{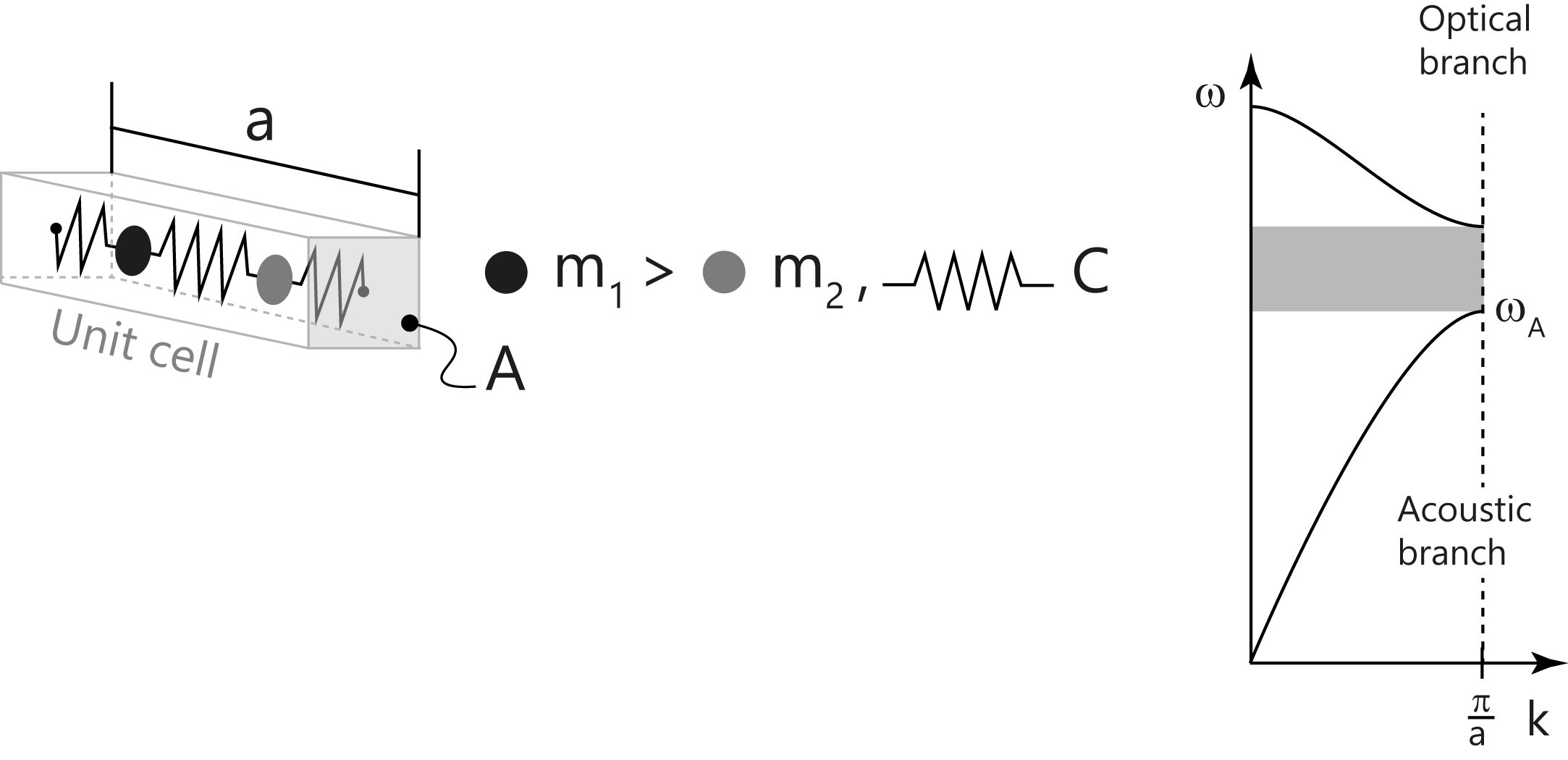}}
\end{center}
\caption{Point-mass PnC.
Unit cell of a diatomic mass-spring chain with volume $V=A\cdot L$, mass elements with mass $m_1 > m_2$ and spring stiffness $C$.}
\label{omegastar}
\end{figure*}

\subsection*{Frequency response of TRO+ and mode shapes}
Figure \ref{fig1} shows a detailed view of the rigid body modes of the TO and TROs. The presence of the anti-resonance in the TRO frequency spectra indicates that an additional resonance can be expected at higher frequency. Indeed, a look at a wider range of frequencies allows us to confirm this expectation, as shown in Fig.~\ref{extfrf}.

\begin{figure*}
\renewcommand{\thefigure}{S\arabic{figure}}
\setcounter{figure}{3}
	\begin{center}
		{\includegraphics[trim=0in 0in 0.in 0in, clip=true, width=.7\textwidth]{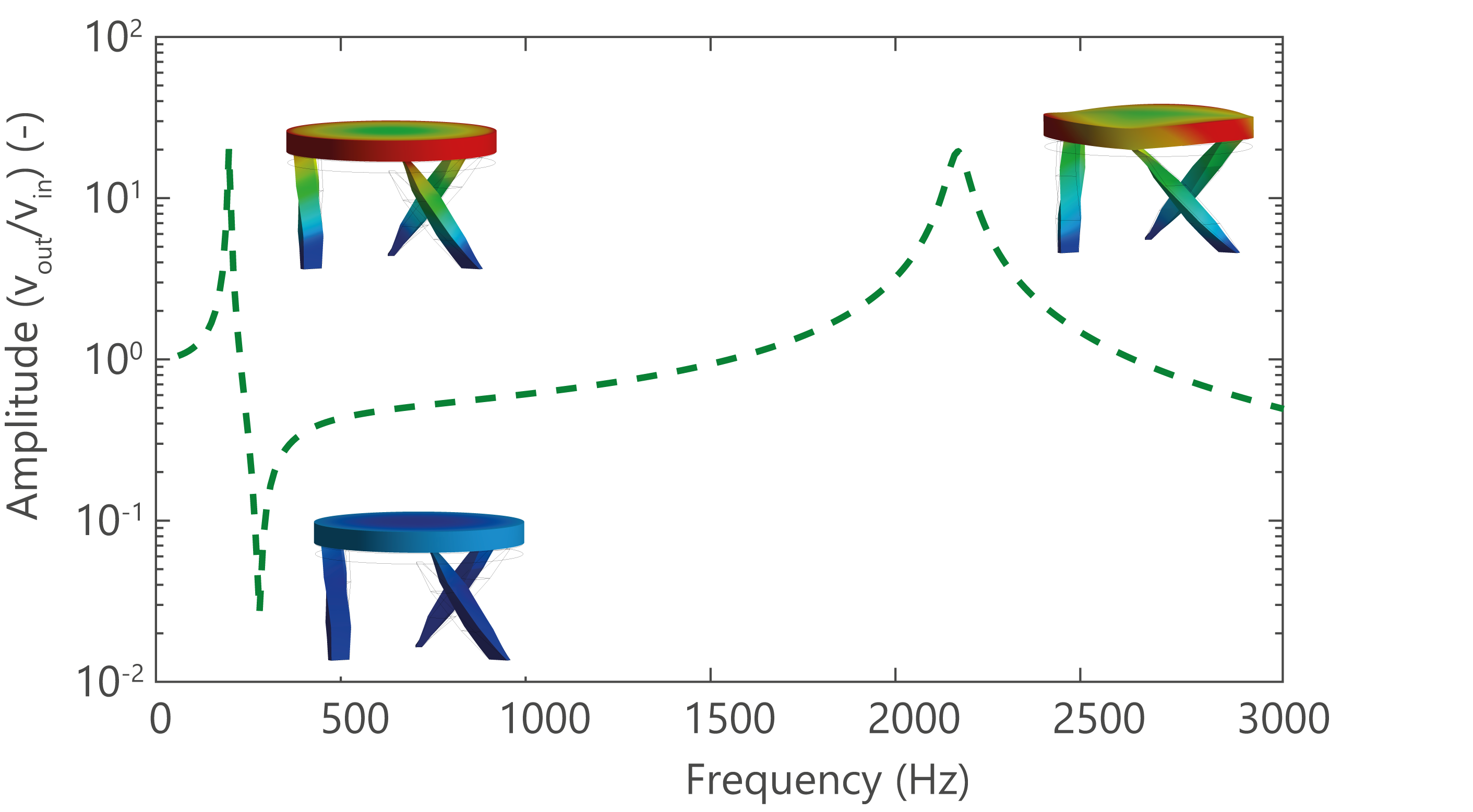}}
	\end{center}
	\caption{TRO+ Frequency response and mode shapes.
	The frequency response of TRO+ shows a further eigenfrequency at approximately 2169 Hz, as can be expected based on the presence of an anti-resonance at about 290 Hz. This is a result of the extended kinematics of the finite size atom oscillator. However, it can also be seen, upon inspection of the mode shapes, that at 2169 Hz, the atom no longer behaves as a rigid body. Animations available online.}
	\label{extfrf}
\end{figure*}
	
\subsection*{Coupling mechanism description}
	
The kinematics and statics of the coupling mechanism between translational and rotary motion of the inertia elements is detailed in Fig.~\ref{fig:analytical}. Here, for a matter of better understanding, the two main functions of the links between atoms are separately represented by the blue spring elements, with stiffness constant $k$ providing a restoring force upon changes of the inter-atomic distance from the equilibrium condition (the analogue of the elastic elements in Brillouin's mass-spring chain) and the black links, representing ideally stiff struts connected to the disks via frictionless pivots, imposing kinematic constraints to the motion of the gray disks by coupling the linear and rotational degrees of freedom.
At the equilibrium condition, the struts connected at a distance $r$ from the axis of the unit cell and laying in a plane tangential to a circle, form an angle $\psi$ to the surface of the disks.
In the geometry presented in Fig.~\ref{fig:analytical}, we consider the syndiotactic unit cell, obtained by mirroring left half with respect to the mid plane $\sigma$ of the middle disk.
The number of struts determines the order of the $n$-fold symmetry axis ($n=3$, in the case at hand).
The kinematics of the isotactic unit cell does not satisfy Cauchy continuum conditions. As such, its properties are determined by the rotational boundary conditions at its ends.

\begin{figure*}
\renewcommand{\thefigure}{S\arabic{figure}}
\setcounter{figure}{4}
\begin{center}
\includegraphics[width=1\columnwidth]{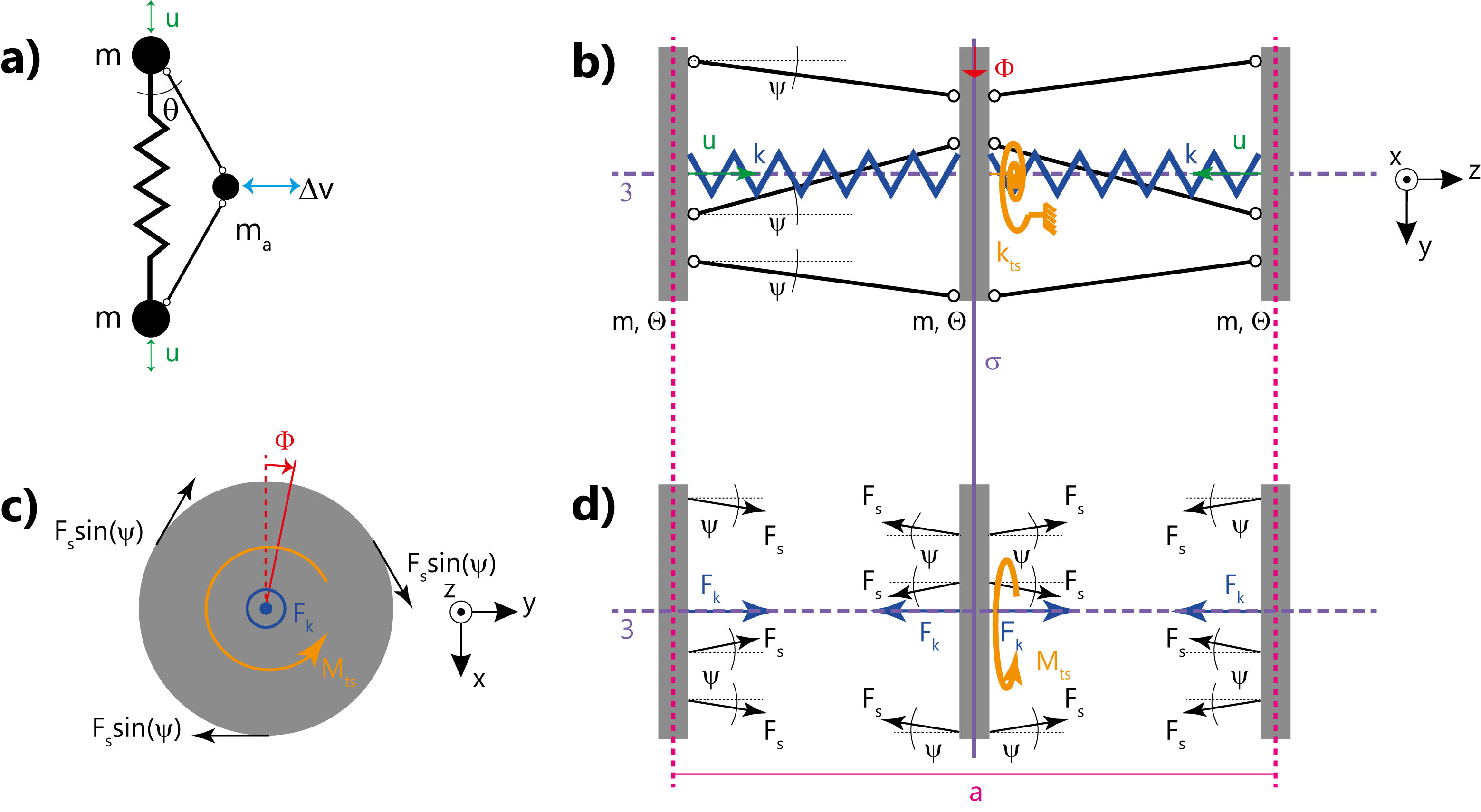} 
\end{center}
\caption{Coupling mechanism.
a), Inertial amplification mechanism proposed in \cite{Yilmaz2007a}, where the range of motion $\Delta v$ of $m_a$ is a multiple of the range of motion $u$ of $m$: $\Delta v=u\cdot \tan\theta$, due to the kinematic constraint imposed by the stiff levers.
b), Example of a PnC with coupled longitudinal rotational degrees of freedom, in which to a displacement $u$ of the outer atoms corresponds a rotation $\Phi$ of the middle atom. This coupling is enforced by the excentric diagonal struts. The elastic elements, a longitudinal and torsional spring, with the stiffness $k$ and $k_{ts}$, respectively, provide the restoring force and moment respectively to return the atoms to the equilibrium position.
c),d), Forces and moments acting on the atoms at equilibrium.}
\label{fig:analytical}
\end{figure*}
	
For small displacements $u$, the rotation $\Phi$ of the middle disk is given by
	
\begin{equation}
\label{eq:kinematic_constraint}
\Phi=\frac{u}{\tan(\psi)}
\end{equation}
	
The restoring force and moment originating from the elements with stiffness $k$ and $k_{ts}$ in longitudinal and rotatory direction are
	
\begin{eqnarray}
\label{eqarr:restoring_forces}
F_k&=&k \cdot u,\\
M_{ts}&=&k_{ts}\cdot \Phi,
\end{eqnarray}

\noindent respectively. From the situation shown in Fig.~\ref{fig:analytical} and the consideration of the kinematic constraints (\ref{eq:kinematic_constraint}), we can write the equation of motion for the PnC as follows
	
\begin{equation}
\label{eq:equation_of_motion}
\ddot{u}\left(m+\frac{\Theta}{\tan^2(\psi) r^2} \right)+u\left(k+ \frac{k_{ts}}{\tan^2(\psi) r^2}\right)=0.
\end{equation}
	
In spite of the fact that part of the energy is coupled into a rotary oscillation, the system described by (\ref{eq:equation_of_motion}) can be still regarded as a one-dimensional mass-spring chain, similar  the one discussed in \cite{Brillouin1946}. Hence, the equation of motion has the form of the equation (\ref{eq:equation_of_motion}), except that the mass terms now include a contribution accounting for the moment of inertia of the disk-shaped atoms, where $\Theta$ for the atoms at hand is given by $\Theta=\frac{1}{2}m r^2$, assuming that the struts are connected to the disk on its circumference. 
	
Therefore, the introduction of chiral elements, made possible by the shape and finite size of the atoms of the PnC, creates additional design elements that can be exploited to tailor the propagation of mechanical waves.
Similarly, the symmetry operations with respect to the atoms of the unit cell represent an additional tailoring property, achievable by tuning the angle $\psi$, which controls the ratio between the longitudinal and rotational motion ($u$ and $\Phi$), and the ratio of moment of inertia $\Theta$ and mass $m$.
As such, we can distinguish in the case at hand between two point groups, namely $\frac{3}{m}$ and $3_1$, as shown in Figs.~\ref{fig2}a,b. 

\subsection*{Additional Modes of Fig.~\ref{fig2}} 
In Fig.~\ref{fig2} additional modes are highlighted as having a strong rotational component in the motion of the atoms (A2 and O2 in Fig.~\ref{fig2}C, M3 and M4 in Fig.~\ref{fig2}D. These additional modes, characterized by some level of local deformation, are reported in Fig.~\ref{fig:extramodes}.

\begin{figure*}[]
\renewcommand{\thefigure}{S\arabic{figure}}
\setcounter{figure}{5}
	\begin{center}
		\includegraphics[width=.5\columnwidth]{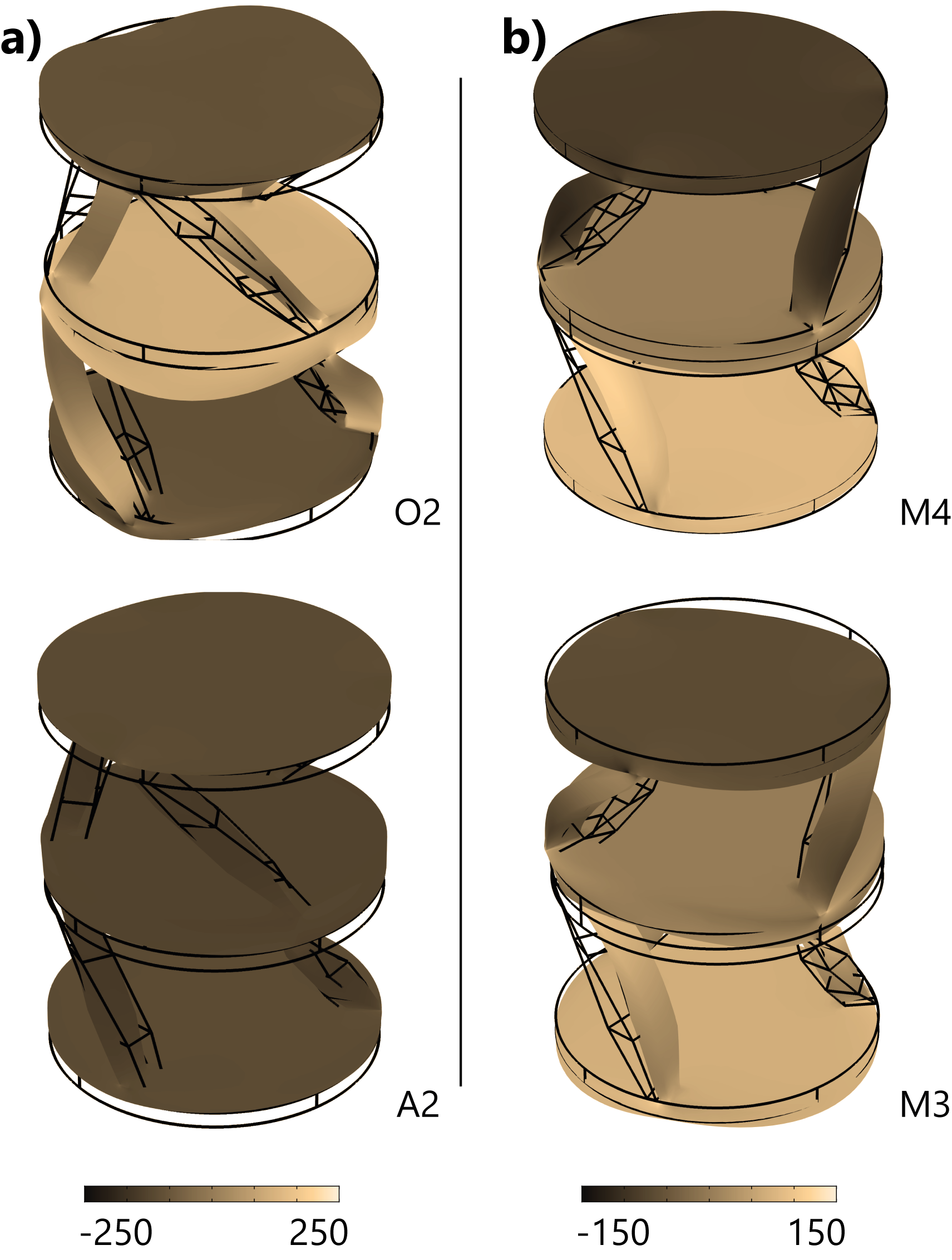} 
	\end{center}
	\caption{Higher modes in figure \ref{fig2} in the iso- and syndiotactic PnCs.
		a), Higher modes of the isotactic PnC: The acoustic (A1) and optical (O1) modes. As for the lower (A1, O1) modes of Fig \ref{fig2}c, also here the behavior is comparable to the one of a mass spring chain, with in-phase motion for the acoustic mode and out of phase motion for the optical one. 
		b), Higher modes of the syndiotactic PnC: M3 and  M4. As for the lower modes M1 and M2 (Fig.~\ref{fig2}d), an unusual behavior (as compared to the isotactic PnC) is observed here, as the top and bottom atoms are always rotating out of phase. In all the modes presented in this figure, we can observe a local component to the deformation of the unit cell, i.e. the disks undergo deformation and do not behave as rigid bodies.}
	\label{fig:extramodes}
\end{figure*}

\subsection*{Data availability}

The data that support the plots within this paper and other findings of this study are available from the corresponding author upon request.

\subsection*{Contributions}
\textbf{A.B.} conceived the research, \textbf{A.B.} and \textbf{M.M.} equally contributed to the writing of the manuscript, the formulation of the concept of tacticity in PnCs and modal analysis of the iso- and syndiotactic unit cells, \textbf{T.D.} introduced the concept of atoms with rotational moment of inertia and contributed initial experiments, \textbf{G.H.} provided finite element models of the PnCs, \textbf{I.L.} contributed the analytical model of the coupling mechanism, \textbf{A.Z.} contributed to the design of experiments and to the writing of the manuscript.

\subsection*{Competing interests }
Certain concepts reported in the present manuscript partially overlap with a patent filed by and granted to Empa (EP 3 239 973). T.D. and A.B. are listed as inventors.

\end{document}